# NMR investigation of atomic and electronic structures of half-Heusler topologically nontrivial semimetals

Chenglong Shi[1], Xuekui Xi[*,1], Zhipeng Hou[1,2], Xiaoming Zhang[1], Guizhou Xu[1], Enke Liu[1], Wenquan Wang[2], Wenhong Wang[1], Jinglan Chen[1], and Guangheng Wu[1]

[1]*State Key Laboratory for Magnetism, Beijing National Laboratory for Condensed Matter Physics, Institute of Physics, Chinese Academy of Sciences, Beijing 100190, People's Republic of China*
[2] *College of Physics, Jilin University, Changchun 130023, People's Republic of China*

ABSTRACT

Recent band structure calculations predict that YPdBi is topologically trivial while its isostructural analogue YPtBi is topologically nontrivial. $^{209}$Bi nuclear magnetic resonance (NMR) spectroscopy is employed to investigate the atomic and electronic structures of both compounds and test this theoretical hypothesis. The observed sign and magnitude of $^{209}$Bi isotropic shifts of YPtBi at various temperatures are systematically distinct from YPdBi. Combined with Hall effect measurements, these results support the band inversion model.

* Author to whom the correspondence should be addressed: e-mail: xi@iphy.ac.cn

## 1. Introduction

There has been a resurgence of interest in the family of RTX (R = rare earth elements including Sc,Y and Lanthanide elements, T=Ni,Pd,Pt, and X=Sb, Bi) half-Heusler alloys, not only for the fundamental importance but also for their potential applications as spintronic or thermoelectric materials [1-3]. Electronic band structure computations have discovered a topologically nontrivial characteristic in terms of a substantial band inversion in some half-Heusler compounds [2-7]. The computed band inversion strength (BIS, defined as energy difference between two bands, $\Delta = E\Gamma_6 - E\Gamma_8$) has been suggested as a measure of spin-orbit coupling (SOC) [4,5]. In particular, first principle computations with appropriate functionals indicate that YPdBi is a topologically trivial narrow band semiconductor/semimetal ($\Delta$>0) while YPtBi is a potential topologically nontrivial semimetal ($\Delta$<0) [4,5,7]. Experimental evidence for such bulk band structural characteristics, however, has rarely been reported [1]. This information should be observable by local susceptibility sensitive methods, such as solid state nuclear magnetic resonance (NMR).

Recently, cubic Yb(Sc,Y)-Pd(Pt)-Sb(Bi) and hexagonal Ce-Pd(Pt)-Sb half-Heusler compounds have been successfully characterized by solid state NMR [8-12]. However, the intrinsic magnetic response of band structural characteristics was frequently masked by localized magnetism from $R^{3+}$ ions. In this work, non 4*f* electron/hole containing YPdBi and YPtBi half-Heusler compounds are investigated to probe atomic and electronic structures by $^{209}$Bi NMR spectroscopy. This choice makes experimental comparison of intrinsic magnetic response of their band structures quite possible since $^{209}$Bi NMR isotropic shift is sensitive to local electronic properties. At room temperature, the measured $^{209}$Bi isotropic shift for YPdBi is (1200±10) ppm while it is (-1748±10) ppm for YPtBi, This large difference is correlated with the theoretically predicted Bi $\Gamma_6$ band inversion and its vicinity to the Fermi energy. Beyond the measurement at room temperature, we've further discovered that $^{209}$Bi NMR shifts are sensitive to temperature in both compounds. In combination with Hall effect measurements and first principle calculations, $^{209}$Bi NMR isotropic shifts at various temperatures are found to reveal more detailed information about their characteristic band structures. This investigation indicates that $^{209}$Bi NMR can be considered as an alternate powerful method to characterize the band structural features that are relevant for understanding their band topology in the family of Bi containing half-Heusler alloys.

## 2. Experimental procedures

Single crystals of YPdBi and YPtBi were grown out of Bi flux and characterized by X-ray diffraction (XRD) with Cu $K_\alpha$ radiation. More details on fabrication procedures can be found elsewhere [13]. $^{209}$Bi NMR spectra of powdered single crystals were obtained by integrating the spin-echo intensity with a Bruker Avance III 400 HD spectrometer in a magnetic field of ~9.39 Tesla. Hahn echo pulse sequence was used for recording echo intensity. The first pulse length (~5 μs) is close to the 90 degree time ($t_{\pi/2}$) of the samples as determined by nutation. A typical pulse delay for the crystals is about 0.3 seconds according to their very short spin-lattice relaxation time (~60 ms). $^{209}$Bi chemical shifts were referenced to 1M $Bi(NO_3)_2$ aqueous nitrate solution ($Bi(NO_3)_3 \cdot 5H_2O$ in nitrate solution at 298 K). The Hall effects were performed by rotating the crystals by 180º in a magnetic field of 5 Tesla using a commercial Physical Properties Measurement System from Quantum Design between 2 and 300 K. Hall coefficients were calculated from the slope of the measured Hall effect curves.

## 3. Results and discussion

Figure 1 presents static $^{209}$Bi (spin $I$ = 9/2) NMR spectra of powdered YPdBi and YPtBi crystals at room temperature. Only central lines can be observed. The central line width defined as full linewidth at half magnitude for YPdBi is ~32 kHz, much larger than that estimated for dipolar broadening (~ 1 kHz), indicating the existence of other contributions such as second order quadrupolar broadening, chemical shift anisotropy or site distributions. The inset in this figure illustrates that this compound crystallizes in a cubic AgAsMg structure with space group number 216 ($F\bar{4}3m$). Y, Bi and Pd atoms occupy the positions 4*a* (0, 0, 0), 4*b* (1/2, 1/2, 1/2), and 4*c* (1/4, 1/4, 1/4), respectively. By the way, the site occupancy in *ref* [13] is incorrect. The electric field gradients (EFGs) at the Bi sites is equal to zero based on crystallographic analysis of point symmetry and first principle calculations. No quadrupole interactions could be expected for ideal stoichiometric phase. Quadrupole interactions show up in the nutation measurements but the magnitudes turn out to be small [14,15], suggesting the existence of some EFGs produced by structural defects and $^{209}$Bi central line broadening from second order quadrupole interactions be small [16]. Different types of atomic disorder are possible including antisite defects in samples. Disorders will cause some EFGs and site distributions. The central line broadening is probably from site distributions. The central line width of YPtBi sample is only ~20 kHz, indicating the site occupancy of Bi atoms in YPtBi are more ordering than YPdBi. In addition, the central lines for both samples are nearly symmetric. This means the anisotropic shifts can be considered to be zero. $^{209}$Bi isotropic shift is then determined by the peak position. This choice can be justified by variable magnetic field dependence of peak position. At room temperature, the measured $^{209}$Bi isotropic shift is about (1200±10) ppm for YPdBi while it is (-1748±10) ppm for its isostructural analogue YPtBi. Similar observation of this large difference in $^{209}$Bi isotropic shifts between the two compounds was also noted in *ref* [12]. The measured shifts and quadrupole parameters including NMR data for YPdSb and YPtSb from literatures [11] are listed in Table 1 for comparison.

$^{209}$Bi NMR isotropic shifts ($^{209}K_{iso}$) and spin-lattice relaxation behaviors of YPdBi and YPtBi were further examined at variable temperatures. In wide band gap semiconducting materials, diamagnetic susceptibility and NMR isotropic shift are usually temperature independent while it is not the case in band gapless semiconductors or semimetals. As shown in Fig. 2a, $^{209}$Bi shifts of YPd(Pt)Bi display temperature dependency. The temperature dependence of $^{209}$Bi NMR shifts for both compounds follows a power law relationship within the same temperature range

investigated. After closer examination, the shifts for YPdBi drop from 1325 ppm to 1170 ppm from 455 to 220 K. For YPtBi, $^{209}$Bi shifts decreases from -1538 to -1777 ppm from 455 to 220 K. Clearly, $^{209}$Bi shifts for YPdBi are systematically distinct from YPtBi, pointing to their distinct electronic structures. The distinct electronic structures between YPdBi and YPtBi can be supported by the temperature dependence of charge carrier concentration (*n*) of YPdBi [13] and YPtBi [17], as determined from the Hall coefficients. Fig.2b shows that *n* for both compounds follows a power law relationship with temperature: $n \propto T^{\alpha}$, where $\alpha$ is a material constant. The fittings yield $\alpha$ = 1.9 for YPtBi and $\alpha$ = 4.0 for YPdBi. This result distinguishes YPtBi from YPdBi and supports the theoretical predication that YPtBi is a topologically nontrivial semimetal since the intrinsic charge carrier concentration of a topologically nontrivial semimetal changes with temperature following the power law, $n \propto T^{3/2}$ [18].

Bulk band structures have been calculated with the Tran-Blaha modified Becke-Johnson density functional [19] implemented in WIEN 2k code [20]. The spin-orbit interactions are respected in the calculations. The radii of muffin-tin (RMT) are set to be 2.5 a.u. for all atoms. The calculated band structures for YPdBi and YPtBi are consistent with previous calculations [4,5]. YPdBi is shown to be trivial while YPtBi is topologically nontrivial, as shown in Fig. 3. The nontrivial topology of the electronic band structure is characterized by the inversion of $\Gamma_6$ band at the $\Gamma$ point. The *s*-like band $\Gamma_6$ in the valence band is situated below *p*-like $\Gamma_8$ and the latter is close to the Fermi level. The band inversion strength $\Delta$ (=$E\Gamma_6$ - $E\Gamma_8$) of YPdBi is 0.29 eV, which is, in magnitude, less than that (-0.87 eV) of YPtBi [4,5]. As shown in Fig. 2a, the observed $^{209}$Bi isotropic shifts of YPtBi at various temperatures are distinct from YPdBi in both sign and magnitude and correlated with the band inversion. This correlation can be simply understood that NMR isotropic shifts of semiconductor or semimetals in a magnetic field are very sensitive to the symmetry of the conduction bands. In YPdBi, the conduction bands have *s*-symmetry while in YPtBi, conduction bands have *p* or *d*-symmetry from transition metals and at the same time Bi Γ6 band is inverted to the valence band due to SOC [2,3]. The *s* character conduction electrons contribute shifts via a Fermi contact mechanism which is positive. The *p or d* character electrons in the conduction band contribute to $^{209}$Bi NMR shifts through negative core polarization mechanism in the materials studied. Again, $^{209}$Bi NMR shifts and spin dynamics in this type of materials are related to the topology of the bulk band structures.

## 4. Conclusions

The atomic and electronic structures of YPdBi and YPtBi have been systematically examined by $^{209}$Bi NMR spectroscopy at various temperatures. The observed $^{209}$Bi NMR shifts and their temperature dependence reveal the distinct nature of band structures between YPdBi and YPtBi. $^{209}$Bi NMR shifts are found to be correlated with the band inversion strength. In combination with Hall effect measurements, our result indicates that $^{209}$Bi NMR isotropic shifts are very sensitive to different chemical environments, which is useful in search of interesting topological nontrivial phases among the family of Bi-containing half-Heusler alloys in the future.

**Acknowledgements** This work was supported by the 973 program (2011CB012800), the NSFC (Nrs. 51371190 and 51171207), and the 100 Talents Program of the Chinese Academy of Sciences.

# Figure Captions

FIG. 1 Static $^{209}$Bi NMR spectra of powdered crystals of YPdBi and YPtBi, taken at 298 K.

FIG. 2 Temperature dependence of $^{209}$Bi NMR isotropic shifts (a) and charge carrier concentration (b) for YPdBi and YPtBi crystals. The lines in (b) are fit with a power law: $n \propto AT^{\alpha}$. The fitting parameter are α=1.9 for YPtBi; α = 4.0 for YPdBi. The dashed lines in (a) are guide for eyes.

FIG. 3 Calculated bulk band structure of YPdBi (a) and YPtBi (b) compounds with spin-orbit interactions. The nontrivial topology of the electronic band structures for YPtBi is characterized by the band inversion of $\Gamma_6$ (red) and $\Gamma_8$ (blue) at the Γ point. The band $\Gamma_6$ symmetry is situated below $\Gamma_8$ which is located at the Fermi level. From orbital symmetries, $\Gamma_6$ is *s*-like band, $\Gamma_8$ is *pd* hybrid bands.

**Table 1** $^{209}Bi$ anisotropic shift, isotropic shift and quadrupole coupling constant, $C_Q$ (= $e^2qQ/h$) parameters for two half-Heusler compounds measured at 298 K.

| Alloys | Anisotropic shift (ppm) | Isotropic shift (ppm) | $e^2qQ/h$ (MHz) | *Ref* |
|---|---|---|---|---|
| YPdBi | 0 | 1200±10 | 0.86 | This work |
| YPtBi | 0 | -1748±10 | ~0 | This work |
| YPdSb | 0 | 924 | ~0 | [11] |
| YPtSb | 0 | 885 | ~0 | [11] |

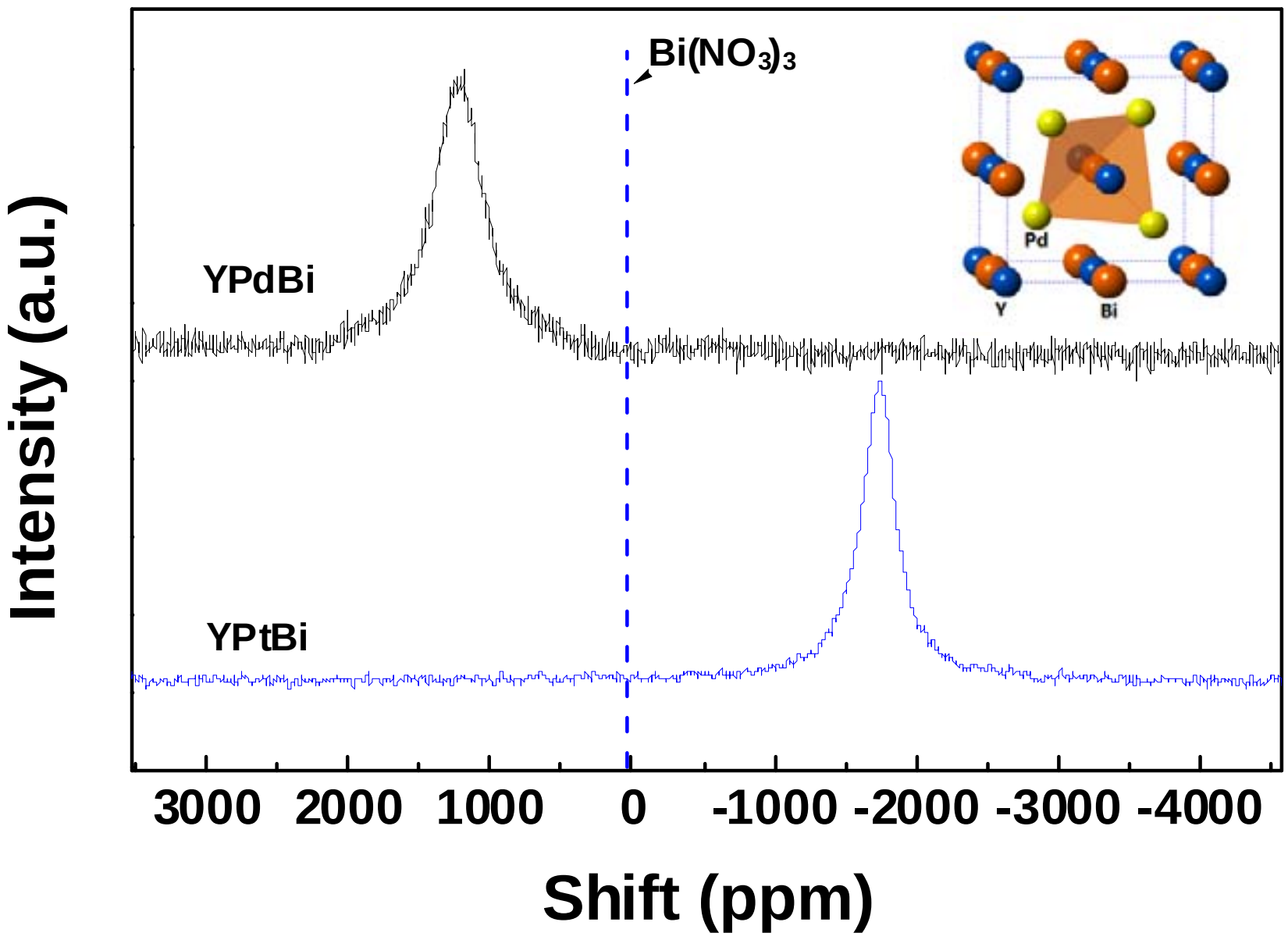

Bi(NO3)3
YPdBi
YPtBi
Pd
Y
Bi
Intensity (a.u.)
3000 2000 1000 0 -1000 -2000 -3000 -4000
Shift (ppm)

**Fig. 1**

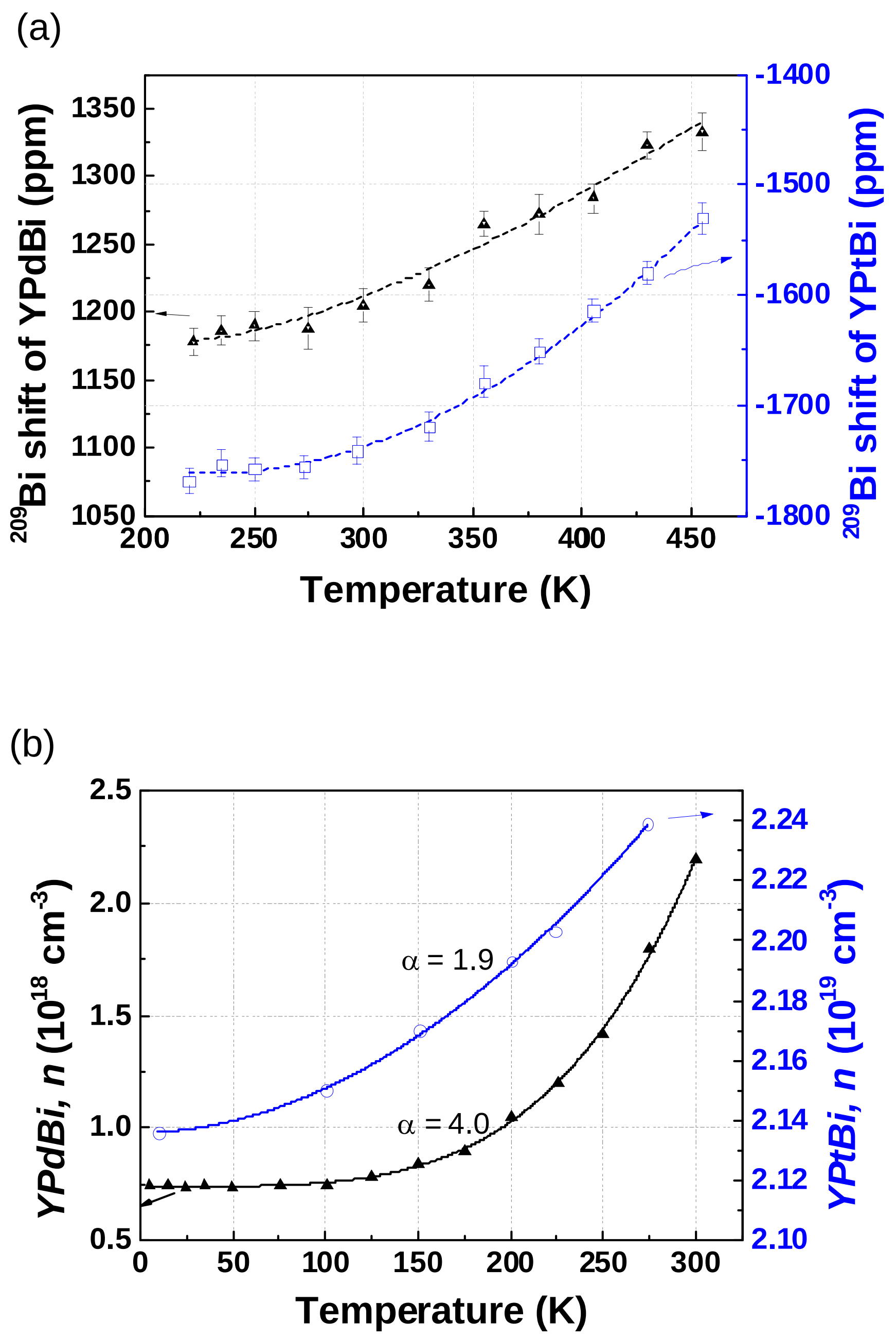

**Fig. 2**

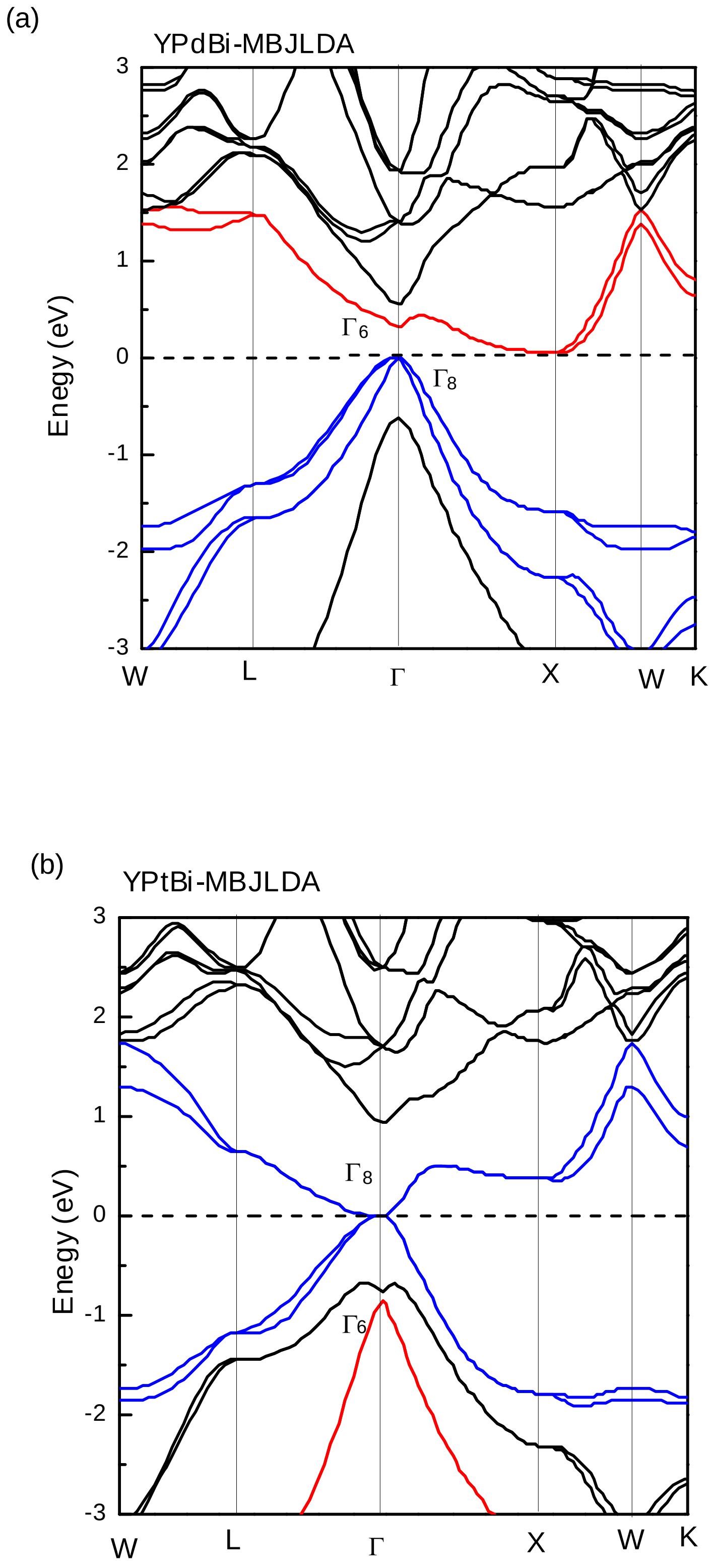

**Fig. 3**